# High-reflectance III-nitride distributed Bragg reflectors grown on Si substrates

M.A. Mastro, R.T. Holm, N.D. Bassim, C.R. Eddy Jr., D.K. Gaskill, R.L. Henry, M.E. Twigg
U.S. Naval Research Laboratory

**Abstract**
Distributed Bragg reflectors (DBRs) composed of an AlN/AlGaN superlattice were grown of Si (111) substrates. The first high-reflectance III-nitride DBR on Si was achieved by growing the DBR directly on the Si substrate to enhance the overall reflectance due to the high index of refraction contrast at the Si/AlN interface. For a 9x DBR, the measured peak reflectance of 96.8% actually exceeded the theoretical value of 96.1%. The AlN/AlGaN superlattice served the added purpose of compensating the large tensile strain developed during the growth of a crack-free 500 nm GaN / 7x DBR / Si structure. This achievement opens the possibility to manufacture high-quality III-nitride optoelectronic devices without optical absorption in the opaque Si substrate.

The III-nitride light emitting diode (LED) market has grown rapidly into a several billion dollar market and will continue to expand as the general indoor/outdoor illumination market develops over the next ten years. To penetrate this market the LED cost/lumen will have to drop by over an order of magnitude [1] and with great certainty the expensive sapphire and SiC substrates will constitute a major fraction of the final packaged LED cost [2]. The obvious choice is to migrate III-nitride epitaxy to low-cost Si substrates that are readily available in large diameters. Nevertheless, epitaxy of GaN on Si is difficult due to the large difference in thermal expansion coefficient and, to a lesser degree, the lattice mismatch with the substrate.

A variety of schemes have been developed to compensate the large tensile stress generated during cool down from growth temperature. Several groups use a superlattice (SL) or graded AlGaN layer inserted near the AlN buffer to introduce a compressive stress during growth that compensates the large tensile thermal contraction stress [3-5]. Nitronex employs an AlN/graded-AlGaN structure to subsequently grow crack-free GaN/AlGaN for the commercial production of microwave transistors [6,7].

Although several groups have demonstrated GaN-based LEDs on Si substrates [8], there are no reports of volume production of GaN LEDs deposited on Si substrates. A major limitation is the opaque Si substrate that absorbs the light emitted from the active region. Insertion of a high-reflectance distributed Bragg reflector (DBR) between the substrate and the active region would increase light extraction by approximately a factor of two, thereby doubling luminous efficiency.

A survey of the literature reveals the previous highest reflectance for a III-nitride DBR on Si was 78% for a 10x AlN/$Al_{0.2}Ga_{0.8}N$ DBR grown by molecular beam epitaxy (MBE) [9]. Most approaches to DBR growth on Si substrates [10] have suffered from the inherent design of the structure. A DBR that does not incorporate the Si substrate as a reflective interface requires approximately 20x quarter-wave stack to achieve 97% reflectance [11]. While this approach is feasible for growth on sapphire or SiC [11,12], the strains generated during growth on Si are prohibitive to such a thick structure.

This paper reports on metal organic chemical vapor deposition (MOCVD) of an AlN/AlGaN SL directly on a Si substrate. The alternating sequence functions optically as a high-reflectance DBR and structurally as a strain compensating SL. Growth was carried out in a modified vertical impinging flow CVD reactor. Two-inch Si (111) wafers were cleaned via a modified RCA process followed by an in-situ $H_2$ bake. An Al seed layer was deposited prior to the onset of $NH_3$ flow to protect the Si surface from nitridation. The AlN/AlGaN SL was deposited at 1050 °C and 50 Torr. In-situ spectroscopic interferometry was used to monitor the layer thickness and reflectance in real time. A fixed wavelength 543.5 nm HeNe laser and an adjustable wavelength source selected from a white light high-intensity lamp are integrated into MOCVD reactors at the Naval Research Lab. The stop-band for the DBR structure can be tuned to any point in the ultraviolet or visible by adjusted the thickness of the AlN and $Al_xGa_{1-x}N$ layers in the DBR. Thus the stop-band of the DBR is adjusted for the device that is subsequently deposited on the DBR. For example, a blue LED with an emission at 458 nm would require a DBR with alternating 55.2 nm AlN and 46.8 nm $Al_{0.05}Ga_{0.95}N$ layers. The GaN cap layer was deposited at 1020 °C at 250 Torr. The higher deposition pressure resulted in GaN layers with larger grains and lower defect densities. Details of this procedure and apparatus have been reported elsewhere [13].

Structural characterization was performed with a Hitachi H-9000 top-entry transmission electron microscope operated at 300 kV and a Panalytical X'pert X-ray diffraction (XRD) system. The ex-situ reflectance was measured at normal incidence using a halogen lamp as a source and reflected beam was dispersed through an Ocean Optics S2000 spectrometer with a 50 µm slit.

Judging from the in-situ reflectance, alloying a small amount of AlN into the GaN altered the growth from a three-dimensional to a two-dimensional mode yielding sharper interfaces and thus improved reflectance despite the decrease in index of refraction contrast. Ng et al. found for AlN/GaN DBR structures on sapphire that the three-dimensional growth mode of GaN produced rough interfaces and thus significantly decreased the overall reflectance of the stack [14]. Although Ng et al. employed MBE as the growth technique, the result indicates the necessity to obtain sharp interfaces via a two-dimensional growth mode. In this work, sharp interfaces are evident in Fig. 1, which is an electron micrograph of a 7x $AlN/Al_{0.05}Ga_{0.95}N$ DBR with a 500 nm GaN cap.

Additionally, the alternating sequence of AlN and $Al_xGa_{1-x}N$ in a SL acts to filter dislocations originating from the interface. Transmission electron microscopy (not shown) found an extremely high level of threading dislocations at the Si/AlN interface (> $10^{12}$ dislocations / $cm^2$). The dislocation level was observed to drop by more than two orders of magnitude through the SL. Details of this mechanism will be reported in more detail elsewhere.

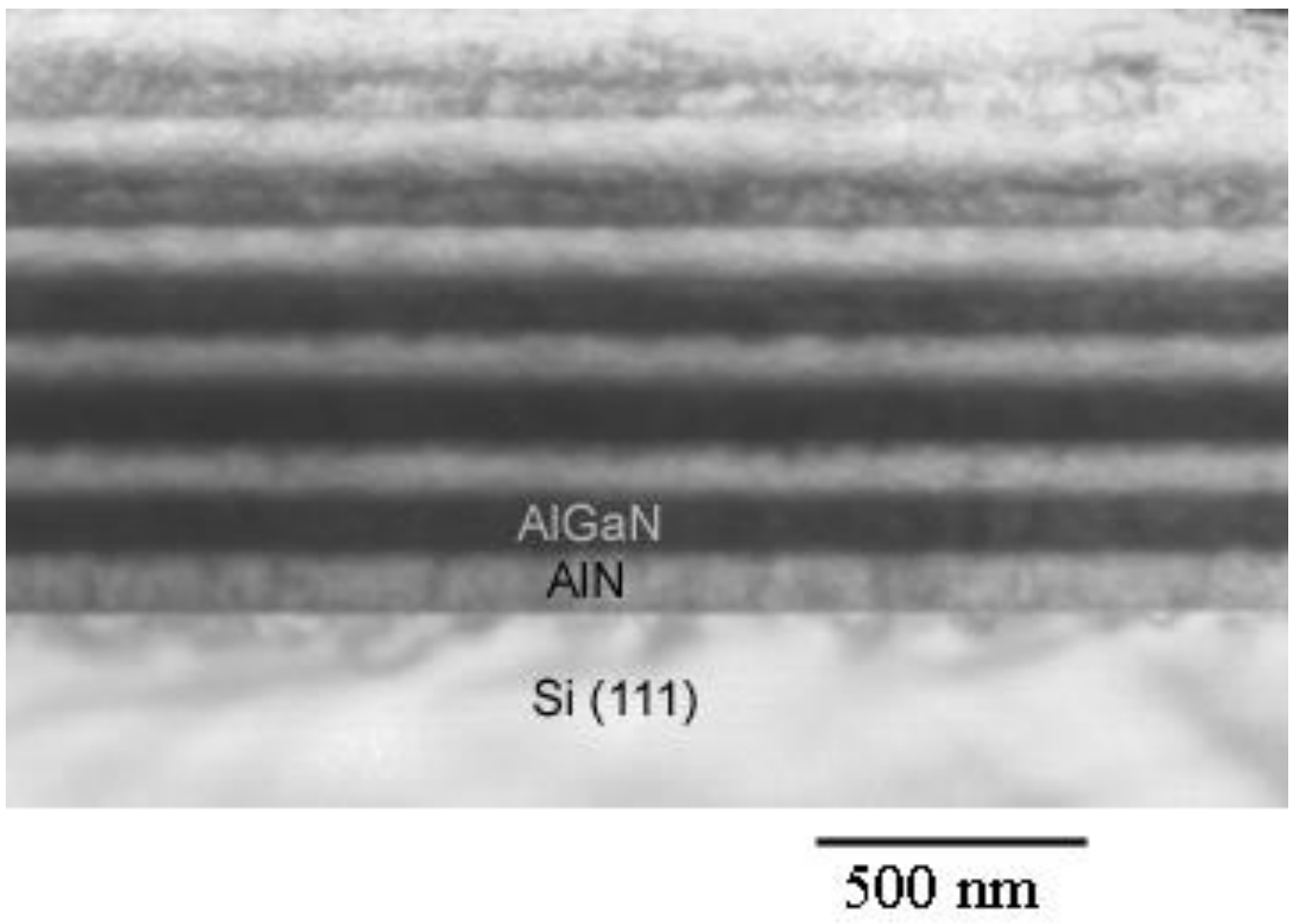

Fig. 1. Dark-field electron micrograph of nominally crack-free 7x $AlN/Al_{0.05}Ga_{0.95}N$ DBR with a 500 nm GaN cap on Si (111) substrate. The interfaces appear to be quite sharp. The decrease in definition at the top of the picture is due an increase in sample thickness and corresponding decrease in electron transparency.

The theoretical reflectance was calculated via the standard transfer matrix method [15] as displayed in Fig. 2. To estimate the reflectance, the calculation assumed an ideal SL for the particular stop-band. A remarkable result for the 9x DBR (Fig. 2a) was that the experimental reflectance of 96.8% actually exceeded the theoretical reflectance of 96.1%. Possibly this difference reveals a small error in the optical constants used in this calculation [16,17]. Regardless, this result indicates the sharpness – in optical dimensions – of the interfaces in the DBR.

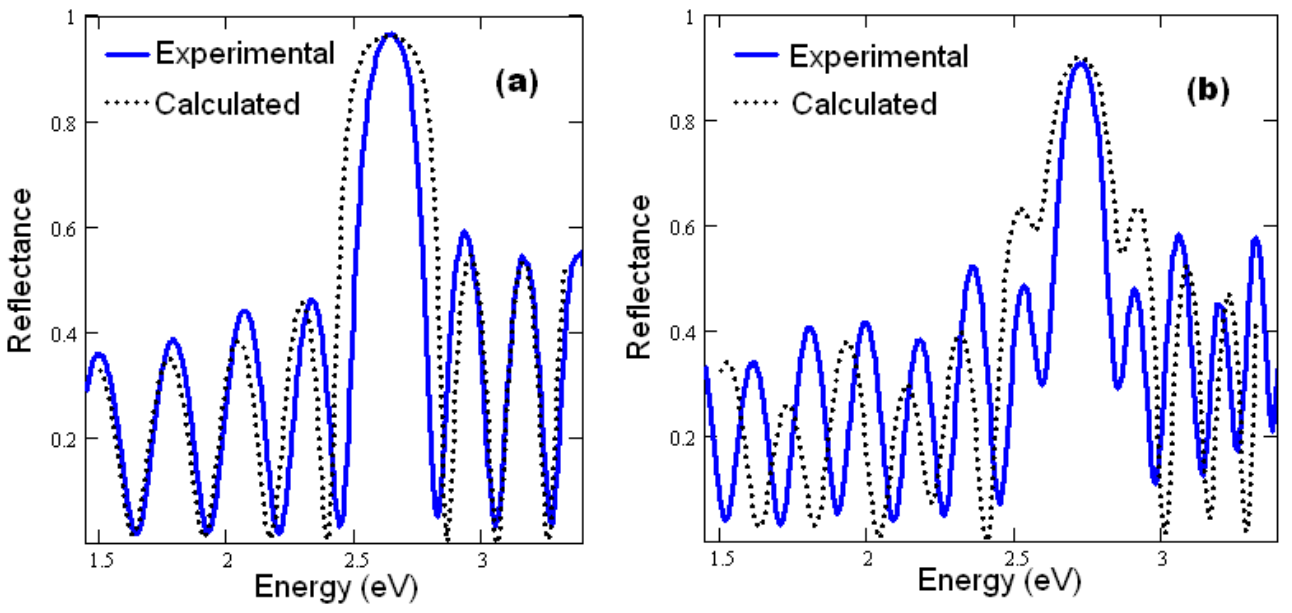

Fig. 2. (Color online). Optical reflectance of a (a) 9x AlN/GaN DBR displaying 96.8% experimental and 96.1% theoretical (for ideal 57 nm AlN / 47.9 nm GaN SL) reflectance with a stop-band centered at λ=472 nm (2.63 eV) and (b) 7x $AlN/Al_{0.05}Ga_{0.95}N$ DBR with a 550 nm GaN cap structure displaying 90.7% experimental and 91.7 theoretical (for ideal 55.2 nm AlN / 46.8 nm $Al_{0.05}Ga_{0.95}N$ SL) reflectance with a stop-band centered at λ=458 nm (2.71 eV)

The wavelength (or energy) dependent refractive indexes are presented in Eq. (1-4). The formulae were fitted to experimental data taken from GaN grown by the authors, AlN from Joo et al. [16], and Si from Palik [17]. The refractive index of $Al_xGa_{1-x}N$ (for x<0.1) was calculated based on a linear relation between the refractive index of GaN and AlN.

$$n_{GaN}(E<3.4eV)=\sqrt{2.89506+\frac{0.11656\lambda_m^{\ 2}}{\lambda_m^{\ 2}-0.1207}+\frac{2.32498\lambda_m^{\ 2}}{\lambda_m^{\ 2}-0.04534}}$$
(1a)

$n_{GaN}(E>3.4eV) = 2.772$ (1b)

$$n_{AlN}=\sqrt{2.396+\frac{1.641\lambda_{nm}^{\ 2}}{\lambda_{nm}^{\ 2}-173^2}} \quad (2)$$

$$n_{Al_xGa_{1-x}N}=xn_{AlN}+(1-x)n_{GaN} \quad (3)$$

$n_{Si}(E>3.32eV) = -20084.501 + 17595.283E - 5.133.409E^2 + 489.912E^3$ (4a)

$n_{Si}(E<3.32eV) = 7.84 - 10.604E + 9.913E^2 - 3.339E^3 - 0.452E^4$ (4b)

where E is the energy in eV, $\lambda_{nm}$ is the wavelength in nanometers and $\lambda_m$ is the wavelength in meters.

In Fig. 3, a strong wurtzite GaN (0002) peak is observable at approximately 34.5° in the 2θ-ω scan. An ω scan (not shown) revealed a FWHM of 0.232°. The FWHM of the GaN in the ω (and the 2θ-ω) scan is exaggerated by the underlying strained AlGaN SL layers whose signal overlaps asymmetrically with the GaN peak. Additionally the AlN (0002) peak is observable but is not clearly defined. Most likely this indicates that the strain state of the AlN varies through the SL, which would smear the XRD spectrum.

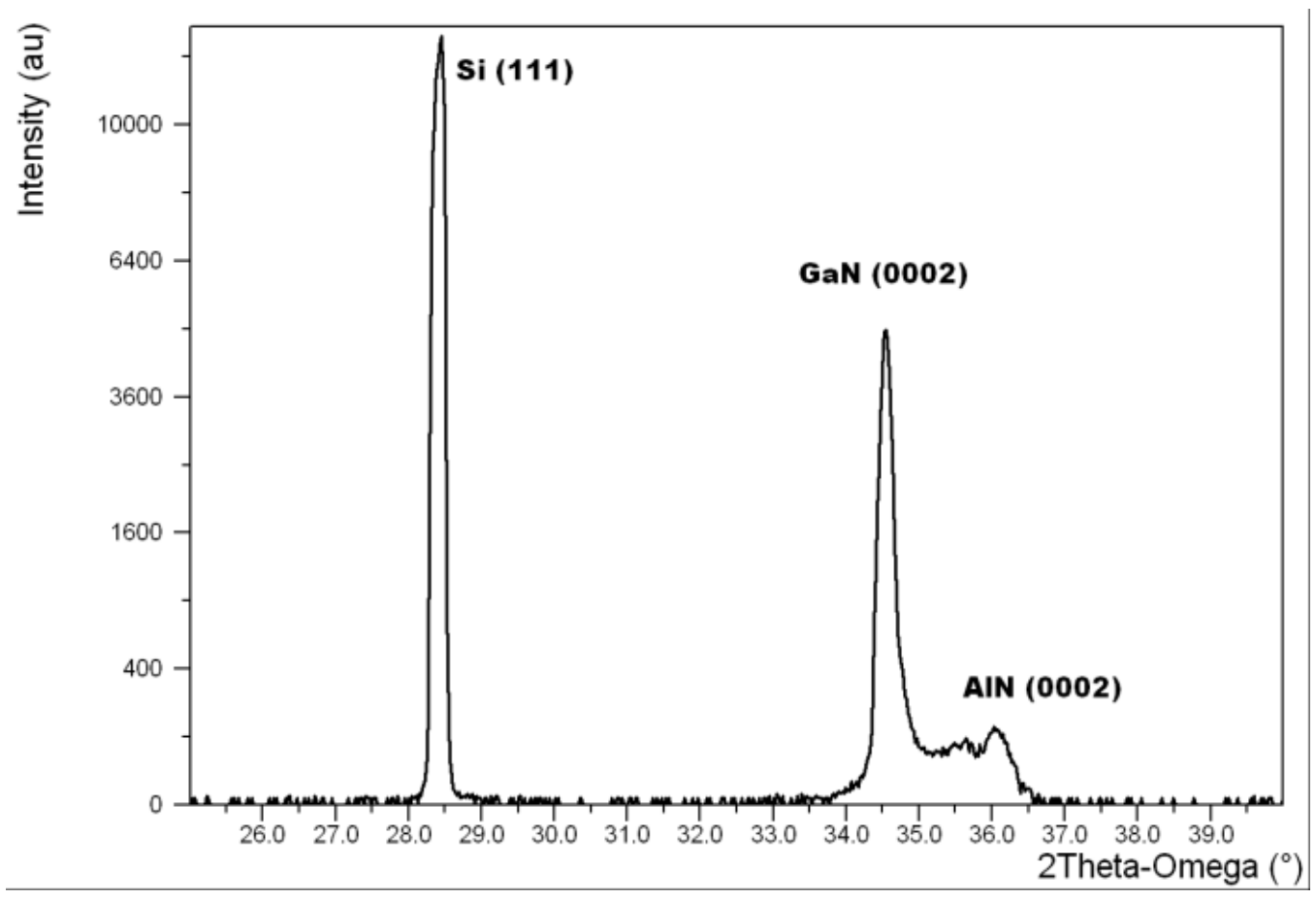

Fig. 3. XRD 2θ-ω scan of 500 nm GaN / 7x $AlN/Al_{0.05}Ga_{0.95}N$ DBR / Si (111) substrate structure

In summary, a high-reflectance (96.8%) III-nitride DBR structure was grown by MOCVD on a Si (111) substrate. The DBR consisted of alternating layers of AlN and GaN (or AlGaN) that introduced a compressive stress to balance the large tensile stress generated during cool down from growth temperature. Additionally, the dislocation density was found to drop by more than two orders of magnitude through a 500 nm / 7x DBR structure. Conceivably, III-nitride LED structures can now be fabricated on low cost Si without optical absorption loss in the opaque substrate.

**Acknowledgements**

Research at the Naval Research Lab is supported by Office of Naval Research; support for Mastro and Bassim was partially provided by the American Society for Engineering Education; the authors thank Mohammad Fatemi and Jim Culbertson for technical discussions

**References**

1. M. Henini, M Razeghi, *Optoelectronic Devices: III Nitrides*, Elsevier Science, Berlin (2005)
2. S. Nakamura, G. Fasol, S.J. Pearton, *The Blue Laser Diode: The Complete Story*, Springer, Berlin (2000)
3. S.-H. Jang, S.-J. Lee, I.-S.Seo, H.-K. Ahn, O.-Y. Lee, J.-Y. Leem, C.-R. Lee, J. Crystal Growth **241**, 289 (2002)
4. A. Reiher, J. Bläsing , A. Dadgar, A. Diez, A. Krost, J. Crystal Growth **248**, 563 (2003)
5. Y. Dikme, G. Gerstenbrandt, A. Alam, H. Kalischa, A. Szymakowski, M. Fieger, R.H. Jansen, M. Heuken, J. Crystal Growth **248**, 578 (2003)
6. Weeks Jr. et al., Patent No: US 6617060 B2; (2003)
7. P. Rajagopal, T. Gehrke, J.C. Roberts, J.D. Brown, T.W. Weeks, E.L. Piner, K.J. Linthicum., MRS Proceedings **743**, L1.2 (2003)
8. W. Weeks and R. Borges, Compound Semiconductor, **7**, 63 (2001)
9. F. Semond, N. Antoine-Vincent, N. Schnell, G. Malpuech, M. Leroux, J. Massies, P. Disseix, J. Leymarie, A. Vasson, Phys. Stat. Sol. (a) **183**, 163 (2001)
10. M. B. Charles, M. J. Kappers C. J .Humphreys, Mater. Res. Soc. Symp. Proc., **831**, E3.17.1 (2005)
11. H. Ng, T. Moustakas, S. Chu, Appl. Phys. Lett., **76**, 2818 (2000)
12. K. Waldrip, J. Han, J. Figiel, Appl. Phys. Lett., **78**, 3205 (2001)
13. C.R. Eddy Jr., R.T. Holm, R.L. Henry, J.C. Culbertson, M.E. Twigg, J. Elec. Mat., **34**(9), 1187 (2005)
14. H. Ng, D. Doppalapudi, E. IIopoulos, T. Moustakas, Appl. Phys. Lett., 74, 1036 (1999)
15. M. Born, E. Wolf, *Principles of Optics*, Pergamon Press (1959)
16. H. Joo, H. Kim, S. Kim, S. Kim, J. Vac. Sci. Technol. A 17, 862 (1999)
17. E. Palik, Handbook of Optical Constants III, Academic Press (1998)